\documentclass[10pt,conference,compsoc]{IEEEtran}

\usepackage{graphicx}
\usepackage{multirow}
\usepackage[T1]{fontenc}
\usepackage{amsmath}
\usepackage{textcomp}
\usepackage{longtable}
\usepackage{url}
\usepackage{colortbl}
\usepackage[margin=1in,footskip=0.25in]{geometry}
\usepackage{breakcites}
\usepackage{afterpage}

\title{Quality-Efficiency Trade-offs in \\ Machine Learning for Text Processing}
\author{\IEEEauthorblockN{Ricardo Baeza-Yates, ~Zeinab Liaghat}
        \IEEEauthorblockA{Web Science and Social Computing Group \\ DTIC, Universitat Pompeu Fabra \\ 
        Barcelona, Catalonia, Spain \\ E-mail: rbaeza@acm.org, zeinab.liaghat@upf.edu}}

\date{}

\begin{document}

\maketitle

\vspace*{0.5\baselineskip}
\begin{abstract}
Nowadays, the amount of available digital documents is rapidly growing from a variety of sources. Extracting information from these documents and finding useful information from such collections has become a challenge, which makes organizing and processing textual big data a necessity. Data mining, machine learning, and natural language processing are powerful techniques that can be used together to deal with this big challenge. Depending on the task or problem at hand, there are many different approaches that can be used. The methods available are continuously being optimized, but not all these methods have been tested and compared in a set of problems that can be solved using supervised machine learning algorithms. The question is what happens to the quality of methods if we increase the training data size from, say, 100 MB to over 1 GB? Moreover, are quality gains worth it when the rate of data processing diminishes? Can we trade quality for time efficiency and recover the quality loss by just being able to process more data? 

We attempt to answer these questions in a general way for text processing tasks, considering the trade-offs involving training data size, learning time, and quality obtained. Hence, we propose a performance trade-off framework and apply it to three important text processing problems: Named Entity Recognition, Sentiment Analysis, and Document Classification. These problems were also chosen because they have different levels of object granularity: words, paragraphs, and documents. For each problem, we selected several supervised machine learning algorithms and we evaluated the trade-offs of these different methods on large publicly available data sets (news, reviews, patents). We use different data subsets of increasing size ranging from 50 MB to several GB, to explore these trade-offs.
For the two last problems, we consider similar algorithms with two different data sets and two different evaluation techniques, to study the impact of the data itself and the evaluation technique on the resulting trade-offs. We find that the results do not change significantly and that most of the time the best algorithms are the ones with fastest processing time. However, we also show that the results for small data (say less than 100 MB) are different from the results for big data and in those cases the best algorithm is much harder to determine. 
%
\end{abstract}

\section{Introduction}

``The challenge is not only to collect and manage vast volumes and different types of data, but also to extract meaningful value from this data'' \cite{Bakshi2012}.  
Indeed, big data makes extracting information a challenge that is both difficult and time consuming. Machine learning (ML) is a powerful tool that can help us with this task. Depending on the task, we need to decide which machine learning algorithm is the most appropriate. One of the relevant criteria, is the ability of the machine learning model to perform accurately on new, unseen examples. Therefore, it is necessary to compare and analyze the result of different models according to different criteria to be able to choose the best possible for each task. However, in practice, this type of comparisons cannot be done as they require extra resources and take more time.

The main goal of natural language processing (NLP) based in machine learning is to obtain a high level of accuracy and efficiency. Unfortunately, obtaining high accuracy often comes at the cost of slow computation \cite{Jiang:2012}. While there is a lot of research to improve accuracy, few consider time and accuracy together, as with big data we need NLP systems to be fast as well as accurate, seeking a reasonable trade-off between speed and accuracy. However, ``what is reasonable for one person might not be reasonable for another'' \cite{Jiang:2012}. The same comment applies to a given task. Hence, we want to find the best algorithm with respect to a customer-specified speed/accuracy trade-off, on a customer proprietary data set. 

Among the supervised machine learning algorithms for a particular task, the algorithms may vary by processing methodology as well as by training efficiency. This makes it difficult for a customer to select an appropriate algorithm for a specific situation. The situation is even more difficult considering that the answer may depend on the specific data set and/or its size as well as the set of algorithms and the type of evaluation used. We can even complicate even more this problem by adding space or time restrictions for the training and/or the prediction phase. In addition, it is hard to find annotated data sets and using professional humans to annotate new training data sets can be expensive and time consuming, even when using crowdsourcing as training data sets can be very large.

With respect to training data size, when it increases, usually quality improves but the algorithm takes more time. However, after a point, quality may not increase as much, while the running time keeps increasing and hence the quality gain may not be worth the efficiency loss. Therefore, increasing training data after that point is not efficient nor effective any more.

We address this problem by considering the trade-offs between training time efficiency and learned accuracy on different sizes of data by studying several algorithms on three different problems/tasks in text processing, comparing them in three dimensions: running time, data size, and accuracy quality. For this we define a framework that allows us to compare different algorithms and define relevant trade-offs. 

The main contributions of this work are the following:
\begin{itemize}
\item A trade-off analysis framework between quality and efficiency that can be applied to most problems that use ML algorithms. In fact, the framework borrows from similar ideas used for generic algorithms.

\item Application of this framework to text processing tasks that are typically solved with supervised ML algorithms, analyzing the impact of the object granularity of the tasks (entities, reviews, documents), the specific data set as well as the type of evaluation used (holdout versus $k$-folds). The main finding is that the best algorithm is not necessarily the one that achieves best quality nor the most efficient one, but the one that balances well both measures for a given training data size. 

\item An experimental comparison of three well-known Named Entity Recognizers (NER) using a news data set that is relevant on its own.
The main result is that the clear winner is the Stanford NER. 

\item An experimental comparison of several ML algorithms for Sentiment Analysis using two subsets of the same reviews data set, to analyze what is the impact of changing the data set when they are of similar type. For one of the subsets we also analyze the impact of the evaluation technique used. The main result is that Support Vector Machines (SVM) is the best algorithm, followed by Logistic Regression (LR), among the algorithms considered.

\item An experimental comparison of several ML algorithms for Document Classification using two different tasks (binary and multi-class) for two different data sets (news and patents), to analyze the impact of changing those parameters. For one of the sets we also analyze the impact of the evaluation technique used.  The main result is that SVM is again the best algorithm, among the ML algorithms considered.
\end{itemize}

Notice that we did not include neural networks in the comparison (that is, deep learning) because their training time complexity is much higher than the most used algorithms and hence they are not competitive in our trade-off analysis. 

The rest of the paper is organized as follows. In Section 2 we present the state of the art, while in Section 3 we explore our problem statement and define our trade-off framework.
In Section 4 we consider the named entity recognition problem while in Section 5 we address sentiment analysis. In Section 6, we address the document classification problems in two different data sets (news and patents). We end with our conclusions in Section 7.
Most of this work is part of the PhD thesis of the second author \cite{Zeinab} and a slightly shorter version of this paper was published in \cite{IEEEBD2017}.

\section{State of the Art}

\begin{figure}[t!]
\centering
\includegraphics[scale=0.5]{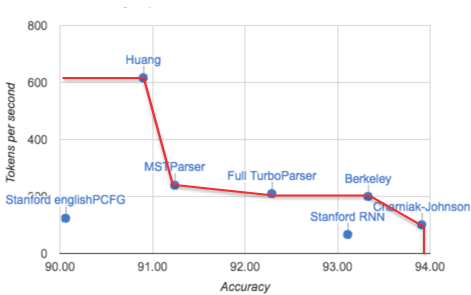}
\vspace*{-0.5cm}
\caption{Accuracy vs. speed for SD parsing \cite{KongS14}.}
\label{fig:fig2_2}
\end{figure}

There are many papers that do experimental comparisons of different ML based algorithms related to text processing, such as \cite{ghandi2012,CerMJM10}, but they usually do not look at the performance trade-off between quality and time.  One exception is Kong and Smith \cite{KongS14} that compares different methods of obtaining Stanford typed dependencies, obtaining the trade-off shown in Figure~\ref{fig:fig2_2}. A similar work on parsing is \cite{Jiang:2012}.


Banko and Brill \cite{Banko:2001} studied natural language disambiguation. They try to find out the effect of training data size on performance and when the benefit from additional training data ends. Ma and Ji \cite{Ma1999} reviewed various general techniques on supervised learning to improve performance and efficiency. They introduce performance as ``the generalization capability of a learning machine on randomly chosen samples that are not included in a training set. Efficiency deals with the complexity of a learning machine in both space and time''. In our case, we are interested in time and quality.


\section{Trade-off Analysis}


As quality is as important as time efficiency, we need a fair way to compare algorithms that achieve different quality at a different processing time cost. Most of the time the best quality algorithm is the slowest one and is not clear if the extra processing time is worth the quality improvement. Hence, we need to explore the trade-offs between quality and time in a way that is independent of the problem being solved as well as the computational infrastructure that is being used. 

Can we trade quality and time and at the end improve both quality and time? Many times, the answer is yes. Let us consider the following example that comes from NER \cite{Baeza-yates_bigdata}.
Let us say that algorithm $A$ finds $\alpha n$ true entities in a text of size $n$ in linear time while algorithm $B$ finds $(\alpha+\varepsilon)n$ true entities in $O(n\log(n))$ time. That is, $B$ has better quality by a margin of $\varepsilon$, but is slower than $A$. However, we are just looking at quality with respect to the data size.

To compare them fairly, we need to consider the same time. So, if both algorithms run in time proportional to $T$, we have that the number of correct entities is:
\begin{eqnarray}
\mbox{A:}& \alpha n  & \in O(T) \nonumber\\
\mbox{B:}& (\alpha+\varepsilon)\frac{n}{\log_2(n)} + O(\log n \log n/\log n) & \in O(T)\nonumber
\end{eqnarray}

Hence, we can equate the two cases to find $n$ such that for some constant $K$, algorithm $A$ finds more correct entities than algorithm $B$, just because it can process more data in the same time, despite achieving less quality. That happens for some $c$ such that:
$$
n > c^{\left(1+\frac{\varepsilon}{\alpha} + O((\log n \log n)/n)\right)}
$$ 
For example, for $\alpha=0.1$ and $\varepsilon=0.05$ with $T(A) = 12 n$ and $T(B) = 2 n \log n$, we find that the point where $A$ starts finding more correct entities than $B$ is when $A$ can process almost 2.4 GB while $B$ can do just over 2 GB.

\subsection{Quality, Time and Data Size}

We can divide a data set into different sizes and measure the training time of different ML algorithms on them. Next, we can calculate the quality of the algorithms using the F-measure \cite{MIR}, as captures sensitivity (recall) as well as specificity (precision). We can use other measures such as accuracy, but usually high accuracy can be achieved by predicting the most common class. Plotting the results, we can find a point on the curve where quality will no longer get better with more data. With increasing data size, obviously, the time increases depending on the algorithm complexity as seen earlier.

\begin{figure}[t!]
\centering
\includegraphics[scale=0.4]{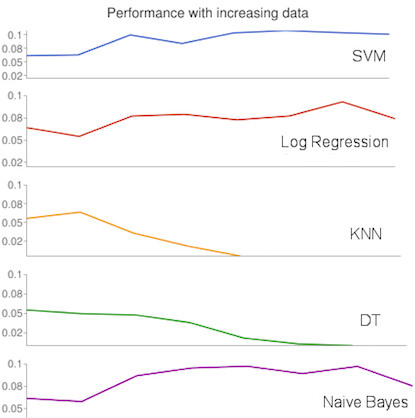}
\caption{Performance vs. training data size example.}
\label{fig:fig3-5}
\end{figure}

To consider the three measures to define our {\em performance} measure as:
\begin{equation}
Performance =  \frac{Quality \cdot Size}{Time}~.   \label{eq3-4}
\end{equation}
That is, performance scales with the size of the data but is penalized by the time consumed. This way, high quality and fast time on large data sets will have very high performance, but high quality with slow time on large data sets will decrease the performance. Figure \ref{fig:fig3-5} is an example of performance change for five well-known classification algorithms applied to data sets of different size. Here we can see that slower algorithms like $k$-Nearest Neighbors (KNN) and Decision Trees (DT) decrease their performance while linear time ML algorithms keep the performance more stable. This performance measure could have different weights in each of the variable considered as well as different functions applied to each variable depending on real costs associated to them.

\subsection{Dominant Algorithms}

An important concept related to trade-offs is a {\em dominant algorithm}. A trade-off graph is a powerful tool for making decisions and usually when one measure improves, another decreases. Dominant algorithms are in the convex hull of all the points in the graph.
In Figure~\ref{fig:fig2_2} we added a red line to show the dominant algorithms for that example.
At one extreme, a choice like Huang would be selecting fast throughput but smaller accuracy. At the other extreme, a choice like Charniak-Johnson would be selecting a high level of accuracy, but slower throughput. According to such graph, an increase of quality involves most of the time a loss in speed. However, we need to avoid choices like Stanford EnglishPCFG or Stanford RNN, that are not the best for any measure. A good choice would be in the frontier, where dominant algorithms are.

Dominant algorithms usually are the same for all data sets, that is, the dependency on data is low. However, this is not always the case.  When we include the number of data sets for which each algorithm is better, the notion of dominant algorithm gets more complicated. We do not use this more complex notion, but a good example for learning to rank algorithms can be found in \cite{Tax2015}.

\subsection{Experimental Design}

Now we explain the experimental rationale that we use in the following three sections for each of the problems that we selected. As we mentioned before, the three problems have different granularity, from words to full documents. The training data sets used also have different sizes due to the availability of large public data sets for each problem. We also use two different evaluation techniques, a simple holdout set for evaluation (20\%) or $k$-fold cross-validation (at least 5 folds).  

In Table~\ref{tab:tab3-12} we show that the rationale is that we start by processing words, then paragraphs and finally full documents. For this reason, we use a news data set for word classification (NER), the Amazon reviews data set for sentiment classification (Movies \& TV and Books, two homogenous subsets for a binary prediction problem), and patents and news for document classification (the first binary and the second multi-class). We also give the maximal data size used and the number of types of features considered. The types of features were basically all the attributes available on the data sets used plus the word vector space, the main type, of the texts considered. In addition, in one case we consider two subsets of the same data set (reviews) and two classes (binary); while in the other case we consider two completely different collections (news and patents) and different prediction problems (multi-class and binary). As we cannot exhaustively compare all the parameters, this selection has a good coverage of all possible combinations. For each problem, we choose several algorithms, where we tried to have two problems with a similar set of algorithms to see the effect of the problem on the results. Another parameter of the experimental space is the number of features, but we decided to keep this parameter of the same order of magnitude for all cases as has a much larger granularity than other parameters being studied and because features do depend on the problem.

\begin{table}[t!]
\centering
\begin{tabular}{|c|c|r|c|}\hline
 {\bf Data set} & {\bf Objects} & {\bf Size~~~} &  { \bf Feature}\\
               &               &               &  {\bf types} \\ \hline\hline
News & Articles (0.8M)  & 2.13 GB & 11 \\
          &  \& words  &                     &           \\ \hline
Movies \& TV  & Reviews (1.8M) & 8.77 GB & 10  \\ \hline
Books  & Reviews (2.6M) & 14.39 GB & 10 \\ \hline
Patents & Documents (7.1M) & 7.18 GB & 6  \\ \hline 
 \end{tabular}
\vspace{.2cm}
\caption{Data sets, object types, sizes, and types of features.}
\label{tab:tab3-12}
\vspace*{-0.5cm}
\end{table}

\begin{table*}[t!]
\centering
\begin{tabular}{|c|c|c|c|c|c|c|c|}\hline
 {\bf Problem (Object)} & {\bf Classes} &  {\bf Data set} & {\bf Algorithms} &  {\bf Evaluation Measures}  & {\bf Evaluation Methods}  \\\hline\hline
NER  (words)  & Multi &  News      &    SNER, LingPipe, Illinois & Precision, Recall, F-measure &  Holdout \\\hline

Sentiment & Binary & Movies \& TV  & DT, LR, KNN, RF, SVM &  Precision, Recall, F-measure & Holdout  \\\cline{2-6}
Analysis   &  Binary  & Books & DT, LR, KNN, RF, SVM &  Precision, Recall, F-measure &  Holdout \&  \\
  ~~~~~(reviews)   &  &        &                     &                & Cross validation (10 folds) \\\hline                     
Document & Multi & News  & DT, LR, KNN, SVM, &  Precision, Recall, F-measure &  Cross validation (6 folds) \\
Classification    &       &       &                   Na\"ive Bayes      &  (micro and macro average) &     \\ \cline{2-6}
 ~~~~~(documents)   & Binary & Patents & DT, LR, KNN, SVM  &  Precision, Recall, F-measure &  Holdout \&  \\       
                 &           &         &          RF       &                   & Cross validation (6 folds) \\\hline                                  
\end{tabular}
\vspace{.2cm}
\caption{Problems, data sets, algorithms, and evaluation measures and methods used.}
\label{tab:tab3-13}
\end{table*}

The characteristics of each problem, algorithms and parameters used are shown in Table \ref{tab:tab3-13}. For Movies \& TV reviews, we used training and testing data sets. For Book reviews and Patents, we use both evaluation methods (holdout as well as cross-validation).  There were no significant differences between holdout or 6 and 10 folds in the cross-validation case. So, we show the results for only one of them in each experiment. 

We used the Scikit-learn framework for the ML algorithms, feature extraction and their evaluation \cite{scikit}.
We use KNN, SVM, LR, DT, Na\"ive Bayes (NB), and Random Forests (RF).
For all the language processing tasks, we used the NLTK toolkit \cite{NLTK}.
For all the experiments, we used a computer with processor Core i7 2.5GHz Intel with $1600$ MHz DDR3 cache CPU and $16$ GB of RAM memory. So our algorithms run almost all the time with the data in main memory.

\subsection{Data Sets}


For news, we used the Reuters Corpus \cite{Lewis:2004} collection, RCV1, that contains about 800 thousand articles in English, taking approximately 2.1 GB when uncompressed. 

With almost $35$ million reviews for almost 2.5 million products from more than 6 million users, Amazon reviews \cite{amazon_dataset} is one of the largest data sets for product reviews. These reviews were collected for $18$ years up to March $2013$. 
The overall data spans over $30$ categories with a total size of about 35 GB. From them we selected the two largest subsets: Books and Movies \& Television. The Mov\&TV subset consists in 2.4M reviews from 70K products, that takes 8.8 GB uncompressed. In the case of the Books subset, it contains 12.8M reviews from almost 930K products, taking 14.4 GB uncompressed. 

Finally, our last data set is derived from the USA Patent databases \cite{patent_dataset}, that 
includes thorough information from patents granted between 1976 and 2016 and patent applications filed between 2001 and 2016, accounting for 7.1M documents that require 7.2 GB of space.


\section{Named Entity Recognition}

We selected the Named Entity Recognition problem because it is one of the main technologies used as a preprocessing step on more advanced NLP applications on different kinds of corpora. We selected for comparison three supervised NER algorithms that are publicly available, well known, free for research, and that are based on machine learning methods:
\begin{itemize}
\item Stanford Named Entity Recognizer (SNER) \cite{Finkel:2005} is based in conditional random fields (CRF);
\item Illinois Named Entity Tagger (INET) \cite{Ratinov2009} uses neural networks (NN) and hidden Markov models (HMM); and
\item LingPipe (LIPI) \cite{lingpipe} that is also based in HMM.
\end{itemize}

\begin{figure}[t!]
\centering
\vspace*{-0.3cm}
\includegraphics[scale=0.45]{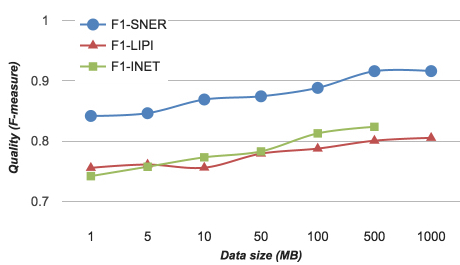}
\vspace*{-0.5cm}
\caption{NER quality comparison vs. data size for News.}
\label{fig:fig4-7}
\end{figure}

The entity types we chose to focus on are Person (PER), Location (LOC) and Organization (ORG), using data subsets from $1$ MB to $1$ GB. 
Figure \ref{fig:fig4-7} shows the overall comparison on quality for the three NERs when used on different data sizes. Our result shows that SNER is clearly better than LIPI and INET in all data sizes. 
In general, with increased data size, quality increases. Once reaching $500$ MB, the quality stabilized and increasing the data did not improve the quality.

\begin{figure}[t!]
\centering
\vspace*{-0.3cm}
\includegraphics[scale=0.45]{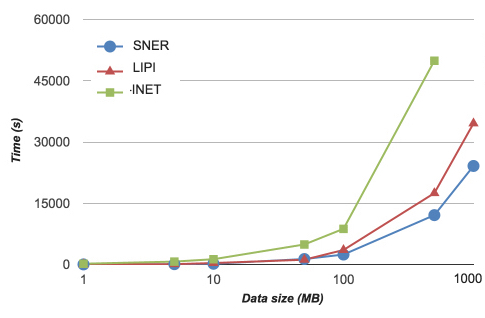}
\vspace*{-0.5cm}
\caption{NER time comparison vs. data size for News (semi-log).}
\label{fig:fig4-8}
\end{figure}

Regarding the running time of these systems in different data sizes, SNER was the best, followed by LIPI and INET for all data sizes as shown in Figure~\ref{fig:fig4-8}.

\begin{figure}[t!]
\centering
\vspace*{-0.3cm}
\includegraphics[scale=0.5]{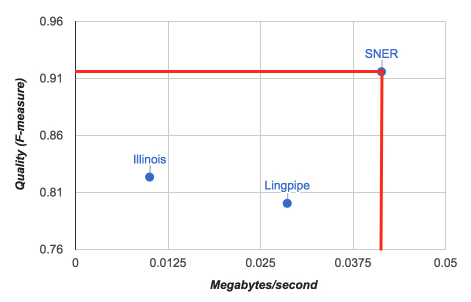}
\vspace*{-0.5cm}
\caption{NER dominant algorithms vs. data size for News (1 GB).}
\label{fig:fig4-10}
\end{figure}

\begin{figure}[t!]
\centering
\includegraphics[scale=0.5]{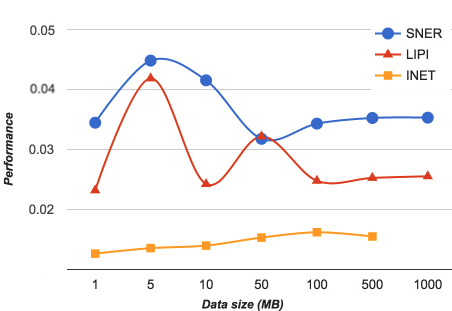}
\vspace*{-0.5cm}
\caption{Performance vs. data size in NER.}
\label{fig:fig4-12}
\end{figure}

Given that SNER has the best quality and is the fastest algorithm, is trivially dominant as is shown in Figure~\ref{fig:fig4-10}.
Now we use our performance measure defined in Equation \ref{eq3-4}, to obtain
Figure \ref{fig:fig4-12}, which shows that overall SNER has better performance, considering all three factors. As the quality does not change after 500 MB, probably the performance will decrease with larger data sets.

\section{Sentiment Analysis}

Sentiment analysis is used for identifying whether a short text is a positive or negative comment and sometimes even the degree of positivity or negativity, such as in web product reviews \cite{wallin2014}. 
The Amazon reviews data set includes the rating of each review, plus the product and user information. The rating is based on one to five stars where one means that the user did not like the product and five means that the user loved the product. So, we used these ratings as training labels (positive for at least four stars and negative for less than three stars). We analyze the same algorithms in two subsets of the Amazon reviews, as explained earlier.

\subsection{Movies and Television Reviews}

\begin{figure}[t!]
\centering
\includegraphics[scale=0.4]{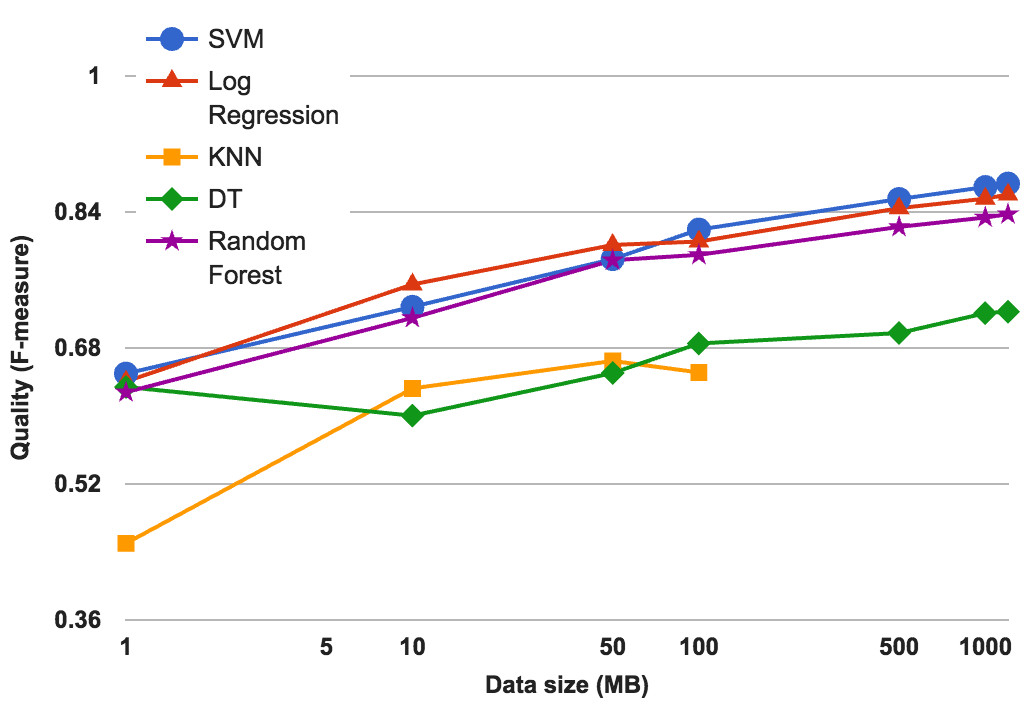}
\vspace*{-0.2cm}
\caption{Quality vs. data size for Movies \& TV reviews.}
\label{fig:fig6-6}
\end{figure}

\begin{figure}[t!]
\centering
\includegraphics[scale=0.4]{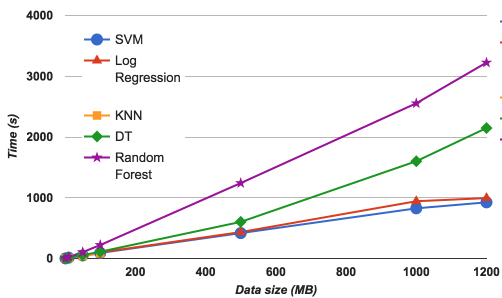}
\vspace*{-0.2cm}
\caption{Running time vs. data size for Movies \& TV reviews.}
\label{fig:fig6-7}
\end{figure}

\begin{figure}[t!]
\centering
\includegraphics[scale=0.4]{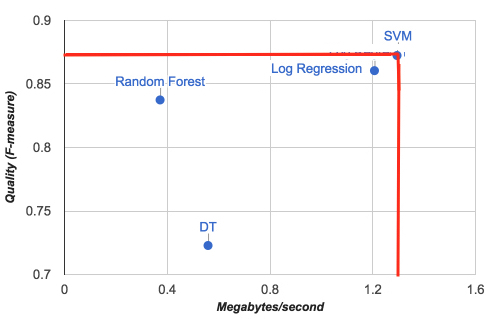}
\vspace*{-0.3cm}
\caption{Dominant algorithms for Movies \& TV reviews (1.2 GB).}
\label{fig:fig6-11}
\end{figure}

\begin{figure}[t!]
\centering
\includegraphics[scale=0.38]{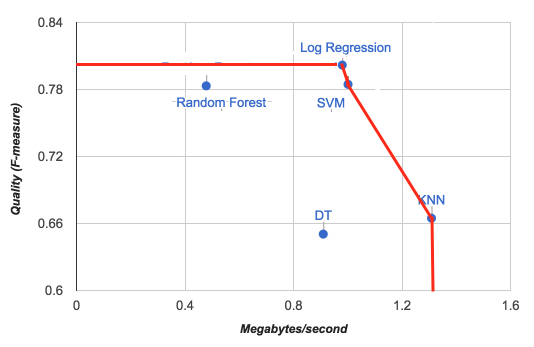}
\vspace*{-0.3cm}
\caption{Dominant algorithms for Movies \& TV reviews (50 MB).}
\label{fig:fig6-11a}
\end{figure}

Figure \ref{fig:fig6-6}  shows the comparison on the quality of the algorithms when run on different data sizes. We see in this case that there is a quality increase when the data size increases. SVM, LR and RF have very close results, but SVM performs better with larger data. LR has the best quality up to 50 MB. Due to the sparsity of the textual data, was very hard to train KNNs on large data. As we can see in Figure \ref{fig:fig6-7}, there is a general linearity for the training time with respect to data size. Because of their inefficiency, selecting RF or DT are poor choices. Figure \ref{fig:fig6-11} shows how SVM becomes the dominant algorithm for big data, as SVM is the fastest in that case. However for small data there are several dominant algorithms as shown in Figure~\ref{fig:fig6-11a}.

\begin{figure}[t!]
\centering
\includegraphics[scale=0.4]{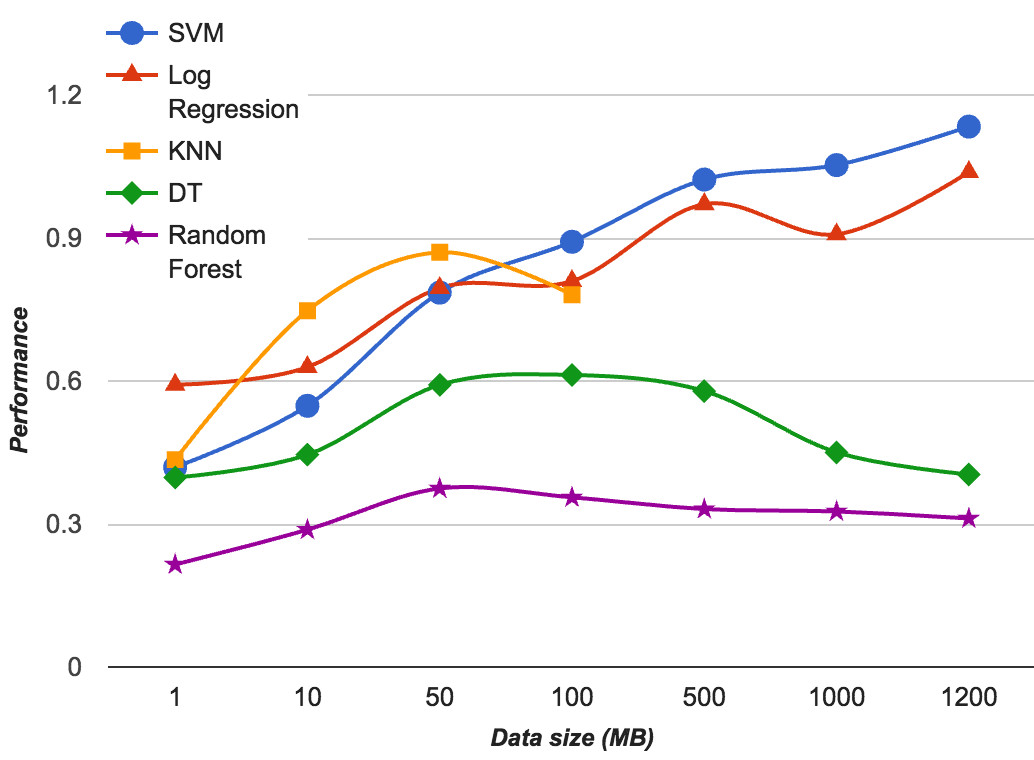}
\vspace*{-0.1cm}
\caption{Performance vs. data size for Movies \& TV reviews.}
\label{fig:fig6-12}
\end{figure}


\subsection{Book Reviews}

\begin{figure}[t!]
\centering
\includegraphics[scale=0.4]{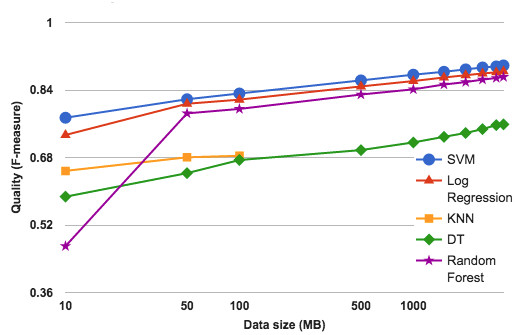}
\vspace*{-.3cm}
\caption{Quality vs. data size for Book reviews.}
\label{fig:fig6-20}
\end{figure}

\begin{figure}[t!]
\centering
\includegraphics[scale=0.4]{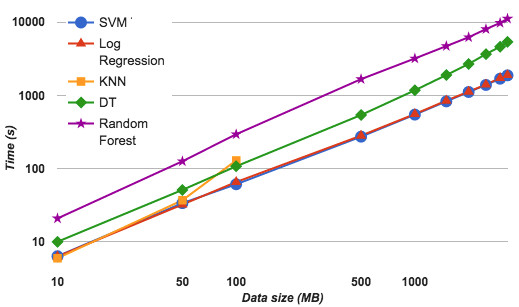}
\vspace*{-.3cm}
\caption{Running time vs. data size for Book reviews.}
\label{fig:fig6-21}
\end{figure}

Figures~\ref{fig:fig6-20} and \ref{fig:fig6-21} compare the different algorithms with respect to quality and time, respectively.
We can see that quality keeps increasing up to the maximal size we had available. If there is a saturation point, happens for 10 GB or more.

\begin{figure}[t!]
\centering
\vspace*{-.3cm}
\includegraphics[scale=0.4]{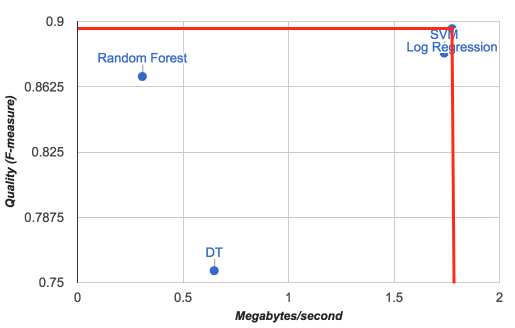}
\vspace*{-.3cm}
\caption{Dominant algorithms for Book reviews (3 GB).}
\label{fig:fig6-24}
\end{figure}

\begin{figure}[t!]
\centering
\vspace*{-.3cm}
\includegraphics[scale=0.21]{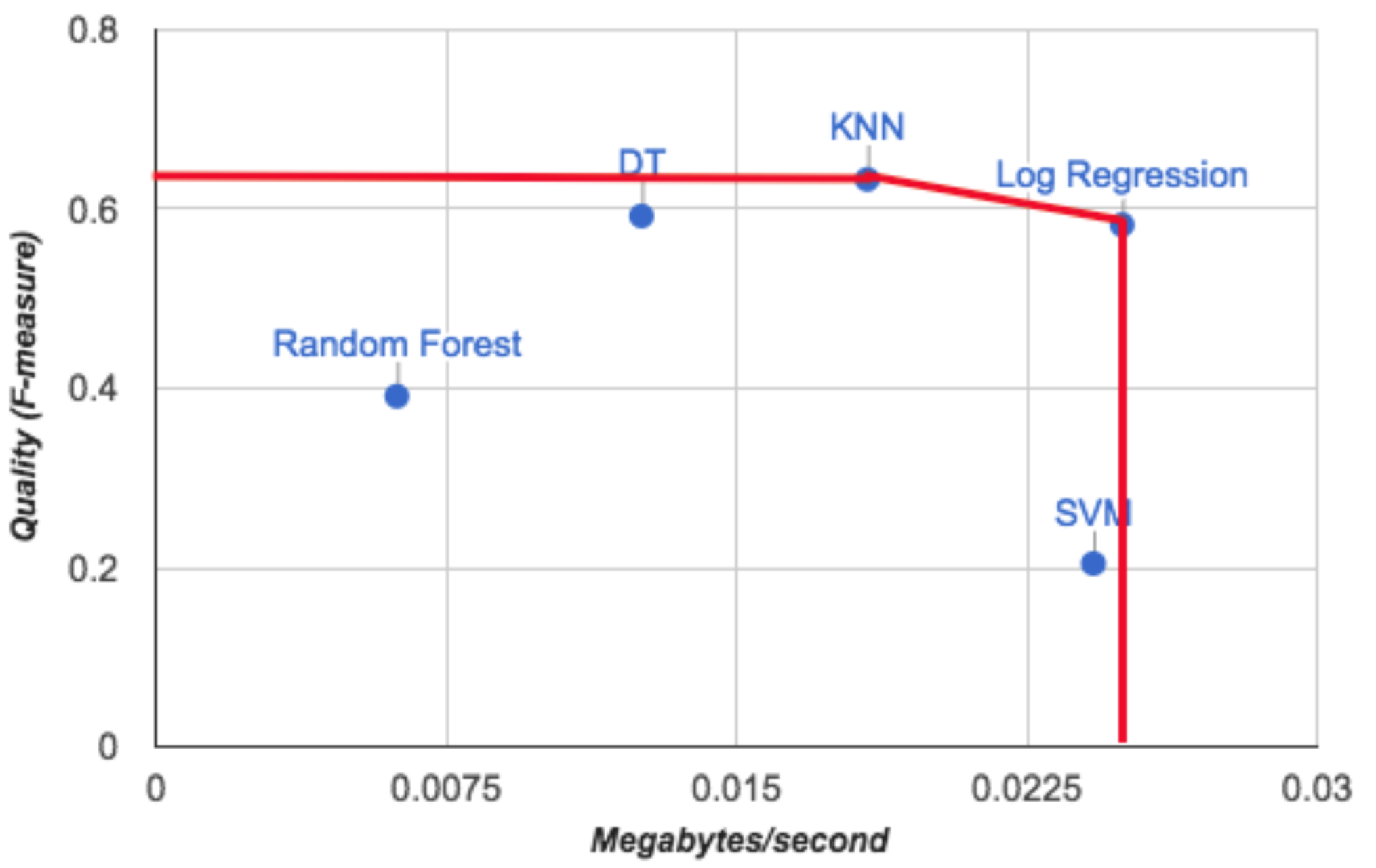}
\vspace*{-.3cm}
\caption{Dominant algorithms for Book reviews (50 MB).}
\label{fig:fig6-24a}
\end{figure}

\begin{figure}[t!]
\centering
\includegraphics[scale=0.35]{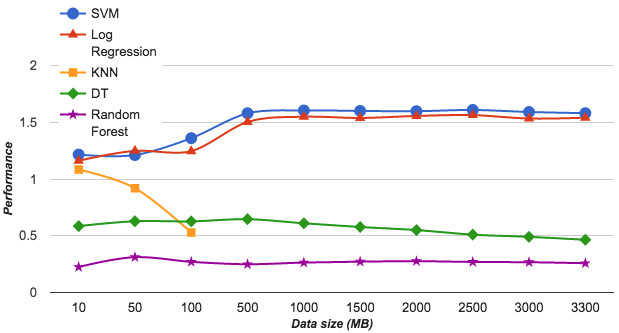}
\vspace*{-.3cm}
\caption{Performance for different data sizes for Book reviews.}
\label{fig:fig6-25}
\end{figure}

For big data, SVM is the dominant algorithm, just beating LR, as shown in Figure~\ref{fig:fig6-24}. However, here again there are
more algorithms for the case of small data as seen in Figure~\ref{fig:fig6-24a}.
Regarding performance, given in Figure \ref{fig:fig6-25}, SVM dominates LR after 10 MB for a small margin, far better than
the other algorithms. DT shows stable performance while KNN and RF performance drops or need too many resources for big data.


\subsection{Discussion}

We tested different algorithms for sentiment analysis with two different subsets of reviews. We got the same result in both experiments, corroborating our intuition that the results should not change when the data is similar. The results show that quality increased by adding more data, and all the algorithms had the same learning curves in quality. However, our results show that in both cases increasing the training data does not always helps to improve the performance. Performance curves were different in both cases, slightly increasing for Movies \& TV reviews and basically constant after 500 MB for Book reviews. 

Considering both cases, if we order the algorithms by quality we have: SVM, LR, RF, and DT for data larger than $100$ MB. By considering only speed (fast-slow), the order changes to: SVM, LR, KNN, DT and RF in all data sizes. Ordering by performance we have: SVM, LR, and KNN for small data. The order for big data (more than $100$ MB) is SVM, LR, DT, and RF. Regarding dominant algorithms, our results show that SVM is the dominant algorithm in data sizes over $50$ MB. KNN, LR, and SVM are dominant in small data.



\section{Document Classification}

The goal of document classification is to automatically assign an appropriate class to each document and is a very popular application \cite{Wajeed2009}. 

The most popular machine learning approaches used in document classification are SVM, DT, KNN, NB, Neural Networks, and Latent Semantic Analysis \cite{ghandi2012,Khan10areview}. In our experiments, we used the first four plus LR, to have a similar set of algorithms to the previous section.


\subsection{News Classification}

\begin{figure}[t!]
\centering
\includegraphics[scale=0.4]{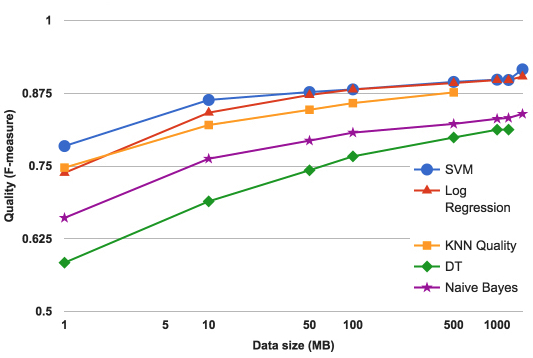}
\caption{Quality vs. data size for News classification.}
\vspace*{-0.3cm}
\label{fig:fig5-13}
\end{figure}

Because we have more than one class, we used macro averaging to compute quality for each class (precision and recall) and then we average all of them to obtain the F-measure. 
In Figure \ref{fig:fig5-13} we show quality with varying training data size for the different algorithms. In this case, we see a saturation effect after 100 MB.

\begin{figure}[t!]
\centering
\includegraphics[scale=0.35]{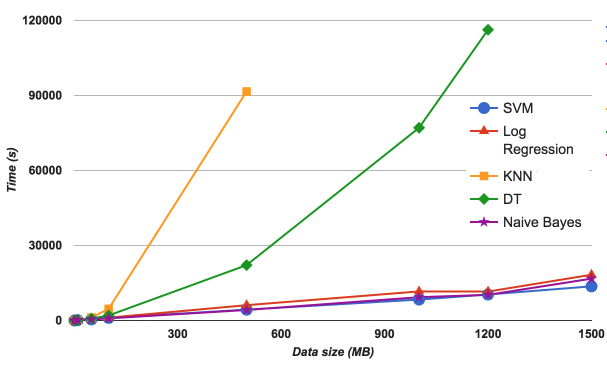}
\vspace*{-0.2cm}
\caption{Time efficiency vs. data size for News classification.}
\label{fig:fig5-14}
\end{figure}

SVM performs better on small data; however, when for larger data sizes LR is equally efficient. 
KNN is near to LR on small and medium data sizes, but as mentioned earlier, KNN is affected by the sparsity of the word vector space. The results for time efficiency are shown in Figure \ref{fig:fig5-14}. Here we can see that SVM, LR, and NB are very similar, while the DTT and KNN are not competitive.

\begin{figure}[t!]
\centering
\includegraphics[scale=0.4]{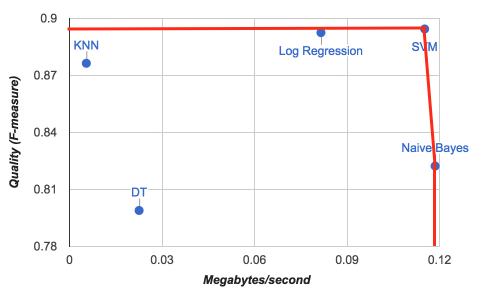}
\vspace*{-.3cm}
\caption{Dominant algorithms for News classification (500 MB).}
\label{fig:fig5-16-2}
\end{figure}

\begin{figure}[t!]
\centering
\includegraphics[scale=0.4]{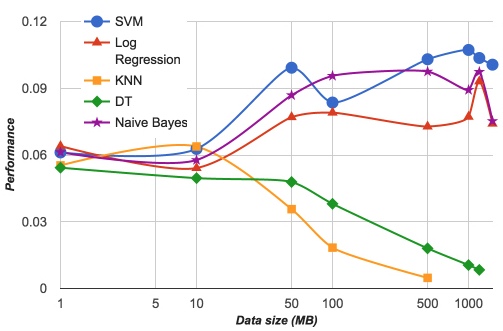}
\vspace*{-.3cm}
\caption{Performance of all algorithms for News classification.}
\label{fig:fig5-17}
\end{figure}

Figure \ref{fig:fig5-16-2} shows the dominant algorithms for 500 MB, one example where SVM and NB are the dominant algorithms up to $1.3$ GB. However, for larger data, SVM is the clear winner and hence is a good algorithm for all data sizes.
The performance comparison is shown in Figure~\ref{fig:fig5-17} where SVM more or less keeps the same performance from 0.5 GB or more.

\subsection{Patent Classification}



\begin{figure}[t!]
\centering
\includegraphics[scale=0.5]{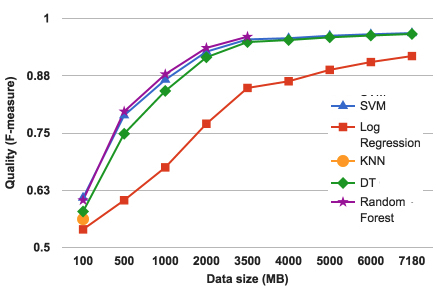}
\vspace*{-0.3cm}
\caption{Quality vs. data size for Patent classification.}
\label{fig:fig7-1}
\end{figure}

\begin{figure}[t!]
\centering
\includegraphics[scale=0.4]{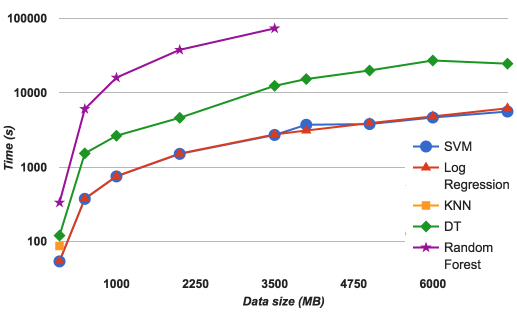}
\vspace*{-0.3cm}
\caption{Time efficiency vs. data size for Patent classification.}
\label{fig:fig7-2}
\end{figure}

Figure \ref{fig:fig7-1} shows the quality comparison of the classification algorithms on various data sizes for the Patents data set.
SVM and DT were the two best algorithms, almost with a perfect tie after 3 GB. 
As Figure \ref{fig:fig7-2} shows, SVM and LR were the fastest algorithms in this case. 

\begin{figure}[t!]
\centering
\vspace*{-0.3cm}
\includegraphics[scale=0.4]{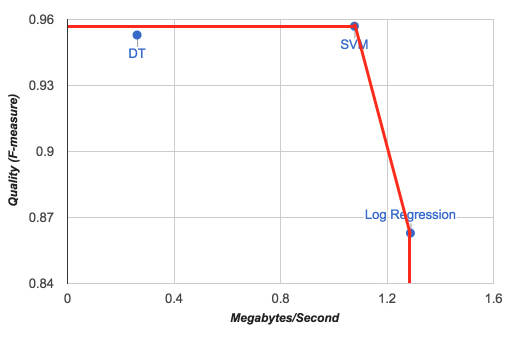}
\vspace*{-0.3cm}
\caption{Dominant algorithms on Patent classification (4 GB).}
\label{fig:fig7-7}
\end{figure}

For all data sizes, SVM dominates. For 1 GB, also RF and LR are dominant. For 4 GB SVM is still dominant, but now together with LR as shown in Figure~\ref{fig:fig7-7}. However, for the largest data set (7.2 GB) only SVM dominates.

\begin{figure}[t!]
\centering
\includegraphics[scale=0.4]{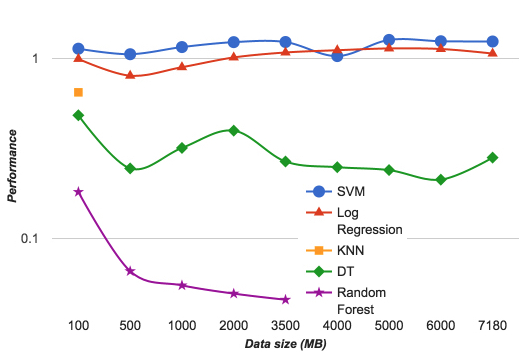}
\vspace*{-0.3cm}
\caption{Performance of all algorithms for Patent classification.}
\label{fig:fig7-8}
\end{figure}

In the overall performance comparison shown in Figure \ref{fig:fig7-8}, SVM wins in all data sizes tested but one, followed closely by LR. Note that the performance is almost the same for all data sizes, something that confirms that the two algorithms have linear complexity.

\subsection{Discussion}

We studied two different document classification problems on different data sets (news and patents). 
In both cases, quality increased by adding more data. All the algorithms had the same learning curves in quality, but the shape of the performance curves was different. 
In addition, algorithms showed different behavior on different data sizes. So, this shows again that adding more data does not always helps to improve the performance.

We also did a $K$-fold evaluation for both cases, news and patents. However, the results did not change with respect to the simple holdout case (80\% training, 20\% for testing).

SVM and LR in News classification and RF and SVM in Patents classification were the best quality algorithms. 
RF and DT show good quality in the latter case because the patent word space is less sparse.
SVM and NB are dominant algorithms in small data, while SVM is dominant in big data.
In both cases, SVM, NB and LR were the fastest algorithms for text classification problems, while KNN and RF were not able to handle large data in our setting. Finally, SVM shows the best performance in larger data sets. 



\section{Conclusions and Future Work}

With the rapid growth of the Web and textual data in general, the task of extracting information from documents in an automatic way, also known as knowledge discovery, has become more important and challenging. Extracting information or knowledge discovery usually uses a combination of ML, NLP, and data mining. A lot of research has been done on this problem using small data sets to try to improve quality. However, few research explores the trade-offs of quality and running time on various data sizes. 
Therefore, our goal was to understand the trade-offs for supervised ML algorithms when dealing with larger data sets.

So, we selected three problems in text processing that are usually solved with supervised machine learning. We compared the performance of several methods based on the quality of the results returned, the training time of the algorithms, and the size of the training data used. We discussed the trade-offs between quality and time efficiency, defining a simple performance measure as quality multiplied by data size and divided by running time. In this way, we can compare fairly different algorithms. We also find the dominant algorithms for each of the problems in different data sizes.
				
The problems considered here included Named Entity Recognition, Sentiment Analysis, and Document Classification. Depending on the problem, kind of data, and number of samples as well as features, the algorithms exhibited different behaviors. For example, SVM had the best performance in larger data sizes. Na\"ive Bayes and KNN showed better performance on small data sizes. However, KNN could not work on large data in our computing setting. 


In conclusion, depending if we focus on quality or speed, we would recommend the use of SVM or Na\"ive Bayes for text classification when the data size is small. Indeed, Na\"ive Bayes requires only a small amount of training data to estimate the constraints parameters necessary for classification. If the number of samples is huge, we would recommend the use of SVM. 
In fact, SVM was the only algorithm that was dominant and had the best performance for the largest data size in all experiments. If the data has many duplicates, DT and RF might be a reasonable choice, as it worked well in data sets which included a significant amount of duplicated data. However, these methods are less efficient. 

Regarding trade-offs, we saw the two possible cases: quality increases with data size or quality goes stable after some point. This is interesting because implies that trade-offs already happen for GB of data. In this case, linear time ML algorithms will at the end be the winners. Indeed, for those algorithms the performance would remain constant independently of data size. Regarding dominant algorithms, our results show that for small data there are several dominant algorithms and hence the decision of which algorithm to use is not trivial.



The first way to extend our work is to do more experiments to cover better the parameter space of the problem of comparing supervised ML algorithms. That implies using more data sets where the notion of dominant algorithm can be extended \cite{Tax2015}, as well as trying all possible evaluation techniques. Another extension would be to vary the number of features and consider more algorithms. We could also add the effect of topic sparsity and text redundancy.


Other future work would be finding the threshold point where quality or performance no longer improves by adding more data for the problems where our data sizes were not large enough. To find the best performance levels, we must be able to estimate the size of the annotated sample required to reach the best performance. 
Another dimension that can be explored is parallel processing and repeat the same experiments using the map-reduce paradigm.

Another extension is to include the prediction time in the trade-off. For example, in many applications, online prediction time could be a constraint that not all ML algorithms may satisfy. This research can also continue by comparing semi-supervised and unsupervised learning methods. All these problems open new interesting trade-off challenges in algorithm analysis and general design for NLP and ML. 



%

\bibliographystyle{IEEEtranS}
\IEEEtriggeratref{8}     
\bibliography{paper}

\end{document}